\documentclass[notitlepage,a4paper,prd,11pt,amsmath,amssymb,amsfonts]{revtex4-1}

\begin{document}
\title{Consistent matter couplings for Plebanski gravity}

\author{Felix Tennie}
\email{felix.tennie@physnet.uni-hamburg.de}
\affiliation{Zentrum f\"ur Mathematische Physik und II. Institut f\"ur Theoretische Physik, Universit\"at Hamburg, Luruper Chaussee 149, 22761 Hamburg, Germany}

\author{Mattias N.\,R. Wohlfarth}
\email{mattias.wohlfarth@desy.de}
\affiliation{Zentrum f\"ur Mathematische Physik und II. Institut f\"ur Theoretische Physik, Universit\"at Hamburg, Luruper Chaussee 149, 22761 Hamburg, Germany}

\begin{abstract}
We develop a scheme for the minimal coupling of all standard types of tensor and spinor field matter to Plebanski gravity. This theory is a geometric reformulation of vacuum general relativity in terms of two-form frames and connection one-forms, and provides a covariant basis for various quantization approaches. Using the spinor formalism we prove the consistency of the newly proposed matter coupling by demonstrating the full equivalence of Plebanski gravity plus matter to Einstein--Cartan gravity. As a byproduct we also show the consistency of some previous suggestions for matter actions.
\end{abstract}
\maketitle
%\tableofcontents

\section{Introduction}
Einstein--Hilbert gravity admits various reformulations~\cite{Peldan:1993hi}. One of the best-known alternative descriptions of the metric geometry of spacetime is Einstein--Cartan gravity~\cite{Cartan} where the fundamental degrees of freedom of the gravitational field are encoded into orthonormal frames and an independent Lorentz connection one-form. While this theory is fully equivalent to general relativity if only tensor field matter is included, it has one major advantage: by construction it allows the immediate coupling of spinor fields. Moreover, from the differential form language used in the formulation of Einstein--Cartan gravity the importance of the two-forms induced by the orthonormal frames was realized. For instance, they can be applied elegantly in the Petrov classification of the curvature of spacetime~\cite{Israel}.  

Plebanski was the first to provide a reformulation of vacuum Einstein--Cartan gravity in which the two-form frames, besides a connection one-form, were considered to be the fundamental variables of the gravitational field~\cite{Plebanski:1977zz}. In this theory, the equations of motion guarantee the existence of an orthonormal frame that induces the two-forms, and from which the metric can then be reconstructed. Plebanski gravity has several interesting features. Its Hamiltonian phase space dynamics~\cite{Peldan:1993hi,Capovilla:1991qb} turns out to be that of the Ashtekar formulation~\cite{Ashtekar:1987gu} of gravity, and so it provides a covariant foundation for the ambitious program of loop quantization~\cite{Thiemann:2007zz}. The Plebanski formulation has also been used as the starting point for the quantization of gravity via spin foam models~\cite{De Pietri:1998mb,Freidel:2007py}. Moreover, it has been argued in~\cite{Krasnov:2006du} that Plebanski gravity could be renormalizable within a certain extended class of theories based on two-form frames.   

In its original formulation Plebanski gravity is a vacuum theory, and in this sense it is an incomplete gravity theory. The specification of matter couplings is essential because it determines the causal propagation of matter fields in spacetime, and in turn the motion of observers. But their precise definition is needed for any interpretation of the gravitational field. 

The matter couplings to Plebanski gravity discussed in the literature are restricted to a perfect fluid model~\cite{Krasnov:2009pu}, and to proposals of actions for scalars, Yang--Mills fields and spinors in~\cite{Capovilla:1991qb}. The energy momentum source terms were argued to be given by two-forms~\cite{Krasnov:2008ui}. Although these results appear comprehensive on a first glance, a closer view shows that they are motivated case by case. There neither exists a general coupling scheme for the standard types of tensor and spinor field matter, nor a general proof of consistency.

In this article we will fill this gap by developing a general minimal coupling scheme for matter fields in Plebanski gravity. We will prove consistency by demonstrating the equivalence of Plebanski gravity with matter to Einstein--Cartan gravity. In section~\ref{sec:EPspin} we will review Einstein--Cartan gravity and the Plebanski formulation in a self-contained way. For this we will explain and use the spinor formalism introduced by Penrose. Section~\ref{sec:ECmat} discusses matter coupling to Einstein--Cartan gravity. In particular we will rewrite the gravitational equations in a form suitable for a later comparison with Plebanski gravity. Section~\ref{sec:matP} contains the essential new results of this article. We will present our new minimal coupling scheme for matter in Plebanski gravity. We will first discuss the simpler case of coupling tensor fields and prove the full equivalence to Einstein--Cartan gravity. With similar technology it is then simple to prove the consistency of some of the known matter couplings, too. Finally we will extend the minimal coupling scheme to include spinor fields. We will conclude with a discussion in section \ref{sec:disc}.

\section{Einstein--Cartan and Plebanski gravity in spinor language}\label{sec:EPspin}
The aim of this section is a review of the Plebanski formalism of gravity which is based on two-form variables that encode the gravitational field. To develop this idea we first discuss Einstein--Cartan theory where the relevant field variables are one-form tetrads. In a second step the tetrads are replaced by two-form frames. In our calculation we employ the spinor techniques introduced by Penrose~\cite{Penrose:1960eq, PenroseRindler}. While these have been used for Plebanski gravity before~\cite{Capovilla:1991qb}, it is hard to find a spinor formulation of Einstein--Cartan theory in the literature.  In section~\ref{sec:matP} the use of spinor methods will become significant in our proof of equivalence of Plebanski gravity with matter and Einstein--Cartan gravity.

\subsection{Einstein--Cartan theory}
In Einstein--Cartan theory, the spacetime metric $g=g_{ab}dx^a\otimes dx^b$ is replaced by a tetrad of one-forms $e^\mu = e^\mu_a dx^a$ with Lorentz indices $\mu=0\dots 3$. Metric and tetrad are related by the orthonormalization condition $g^{-1}(e^\mu,e^\nu)=g^{ab}e^\mu_a e^\nu_b=\eta^{\mu\nu}$ with canonical Lorentzian metric $\eta^{\mu\nu}=\mathrm{diag}(1,-1,-1,-1)^{\mu\nu}$. In the first-order formulation the vacuum action for the tetrad fields employs an $SO(1,3)$-connection one-form ${\omega^{\mu\nu}=\omega^{\mu\nu}_a dx^a=-\omega^{\nu\mu}}$, and reads
\begin{equation}\label{eqn:ECTaction}
\tilde S^\mathrm{EC}[e^\mu,\omega^{\mu\nu}] = \int_M \epsilon_{\mu\nu\rho\sigma}\Big(e^\mu\wedge e^\nu\wedge R^{\rho\sigma}[\omega] - \frac{\lambda}{3}e^\mu\wedge e^\nu\wedge e^\rho\wedge e^\sigma\Big).
\end{equation}
Here $R^{\rho\sigma}[\omega]=d\omega^{\rho\sigma}+\omega^\rho{}_\tau\wedge\omega^{\tau\sigma}$ denotes the curvature two-form of $\omega$ (which should not be confused with the Ricci tensor), and $\epsilon_{\mu\nu\rho\sigma}$ is the totally antisymmetric symbol with $\epsilon_{0123}=1$. Variation of the action $\tilde S^\mathrm{EC}$ with respect to $\omega$ yields
\begin{equation}\label{eqn:ECT1}
de^\mu+\omega^\mu{}_\nu\wedge e^\nu=0\,,
\end{equation}
which is solved by the torsion-free Levi-Civita connection $\omega[e]$. The dynamical equations for the tetrad fields are obtained by independent variation and read
\begin{equation}\label{eqn:ECT2}
\epsilon_{\mu\nu\rho\sigma}\Big(e^\nu\wedge R^{\rho\sigma}[\omega] - \frac{2\lambda}{3}e^\nu\wedge e^\rho\wedge e^\sigma\Big)=0\,.
\end{equation}
Substituting the solution $\omega[e]$ these have been shown to be equivalent to vacuum Einstein gravity with cosmological constant $\lambda$~\cite{Trautman:2006fp}.

We will now present a reformulation of the Einstein--Cartan action $\tilde S^\mathrm{EC}$ in the spinor formalism. Observe that all objects appearing in the action~(\ref{eqn:ECTaction}) are Lorentz algebra valued tensor fields. These will be replaced by spinor fields, which works as follows. Lorentz tensor fields are sections of tensor bundles associated to an underlying $SO(1,3)$ principal fiber bundle. Since  $SL(2,\mathbb{C})$ is the universal covering group of $SO(1,3)$, one may instead consider an $SL(2,\mathbb{C})$ principal fiber bundle. Then one constructs the associated tensor bundles where tensors now are allowed to be multi(anti-)linear maps (using complex conjugation). Sections of these associated bundles are called spinor fields and form the spinor algebra. It turns out that the subalgebra of Hermitian spinor fields is isomorphic to the algebra of Lorentz tensor fields. An explicit map between these is provided by the Infeld--van der Waerden symbols~$\sigma$,
\begin{equation} 
T^{\alpha\dots}{}_{\beta\dots} \mapsto T^{AA'\dots}{}_{BB'\dots}=\sigma^{AA'}_\alpha\dots\sigma^\beta_{BB'}\dots T^{\alpha\dots}{}_{\beta\dots}\,.
\end{equation}
The capital Latin indices $A,A'$ may take values $0,1$. Up to normalization, $\sigma^{AA'}_0$ is the $2\times 2$ identity matrix while $\sigma^{AA'}_1\dots \sigma^{AA'}_3$ are the Pauli matrices. The Infeld--van der Waerden symbols satisfy the inversion relations
\begin{equation}
\sigma^{AA'}_\mu\sigma^\mu_{BB'} = \delta^A_B\delta^{A'}_{B'}\,,\qquad 
\sigma^\alpha_{MM'}\sigma^{MM'}_\beta = \delta^\alpha_\beta\,,
\end{equation}
which implies that the Lorentz indices in  fully contracted Lorentz tensor expressions can simply be replaced by pairs of spinor indices. This also holds for densities like the Lagrangian density of~$\tilde S^\mathrm{EC}$. 

We now discuss the spinor equivalents of the action ingredients in turn. The tetrad fields and the connection are mapped into Hermitian spinor fields as $e^\mu\mapsto e^{MM'}$ and $\omega_{\mu\nu}\mapsto \omega_{MNM'N'}$; note that it is only the order within primed or unprimed indices that plays a role. Due to its antisymmetry, $\omega_{MNM'N'}=-\omega_{NMN'M'}$, the $SO(1,3)$ connection can be decomposed in the form
\begin{equation}\label{eqn:conndec}
\omega_{MNM'N'} = \omega_{MN}\epsilon_{M'N'} + \bar\omega_{M'N'}\epsilon_{MN}\,,
\end{equation}
where $\epsilon_{MN}=\epsilon^{MN}$ (with primed or unprimed indices) are totally antisymmetric symbols with $\epsilon_{01}=\epsilon^{01}=1$. These matrices $\epsilon$ are used to raise or lower spinor indices according to the so-called northwest--southeast convention, e.g., $\omega^P{}_N = \epsilon^{PQ}\omega_{QN}$ or $e_A{}^{C'} = e^{PC'}\epsilon_{PA}$. Note that
\begin{equation}\label{eqn:edel}
\epsilon_B{}^A = \delta^A_B = -\epsilon^A{}_B\,.
\end{equation}
The components $\omega_{MN}$ and $\bar\omega_{M'N'}$ in (\ref{eqn:conndec}) are completely determined by the original components $\omega_{MNM'N'}$. A similar decomposition exists for the curvature $R_{\mu\nu}\mapsto R_{MNM'N'}$ which is also antisymmetric,
\begin{equation}\label{eqn:curvdec}
R_{MNM'N'}  = R_{MN}\epsilon_{M'N'} + \bar R_{M'N'}\epsilon_{MN}\,.
\end{equation}
One may check that  
\begin{equation} 
R_{MN} = d\omega_{MN} + \omega_{M}{}^P \wedge \omega_{PN}\,.
\end{equation}
The one-forms $\omega_{MN}$ form an $SL(2,\mathbb{C})$ connection with associated curvature two-forms $R_{MN}$. Final ingredient of the action~(\ref{eqn:ECTaction}) is the epsilon symbol $\epsilon_{\mu\nu\rho\sigma}$ that is mapped to 
\begin{equation}
\epsilon_{MNPQM'N'P'Q'} = i \left(\epsilon_{MP}\epsilon_{NQ}\epsilon_{M'Q'}\epsilon_{N'P'}-\epsilon_{MQ}\epsilon_{NP}\epsilon_{M'P'}\epsilon_{N'Q'}\right).
\end{equation}

Inserting the spinor equivalents of the tetrads, connection and curvature, and of the epsilon symbol into the action~(\ref{eqn:ECTaction}) yields the spinor formulation of Einstein--Cartan gravity:
\begin{eqnarray}\label{eqn:ECTact2}
S^\mathrm{EC}[e^{MM'},\omega_{MN},\bar\omega_{M'N'}] &=&  -2i \int_M e^{AA'}\wedge e^{BB'} \wedge\left(R_{AB}[\omega]\epsilon_{A'B'} - \bar R_{A'B'}[\bar\omega]\epsilon_{AB} \right)\nonumber\\
&& \qquad \qquad- \frac{\lambda}{3} e^{AA'}\wedge e^B{}_{A'}\wedge e_A{}^{C'}\wedge e_{BC'}\,.
\end{eqnarray}
Note that the imaginary factor in front of the integral ensures that the action is real. The equations of motion are derived by variation with respect to the dynamical variables $\omega_{MN}$ and $e^{MM'}$:
\begin{subequations}\label{eqn:012} 
\begin{eqnarray}
\!\!\!\!\!  d\left(e^{MA'}\wedge e^N{}_{A'}\right) - 2 \omega^{(M}{}_A\wedge e^{N)A'}\wedge e^A{}_{A'}& = & 0\,,\label{eqn:spin1}\\
e^{AA'} \!\wedge \!\left( R_{MA}[\omega]\epsilon_{M'A'} - \bar R_{M'A'}[\bar\omega]\epsilon_{MA}\right) &=& \frac{2\lambda}{3} e^A{}_{M'}\wedge e_M{}^{A'}\wedge e_{AA'}\,.\label{eqn:spin2}
\end{eqnarray}
\end{subequations}
The equation obtained by variation with respect to $\bar\omega_{M'N'}$ is found to be the complex conjugate of the equation of motion for $\omega_{MN}$. 

We will now show the equivalence of the equations of motion (\ref{eqn:012}) of the spinor formulation to the original Lorentz tensor field version of Einstein--Cartan gravity with equations (\ref{eqn:ECT1}) and (\ref{eqn:ECT2}). Using the Infeld--van der Waerden map, equation (\ref{eqn:ECT2}) can be rewritten in terms of spinorial objects. The appearing curvatures can then be decomposed as in (\ref{eqn:curvdec}) which yields equation (\ref {eqn:spin2}). Both these steps are equivalent transformations. 

Proving the equivalence of equations (\ref{eqn:ECT1}) and (\ref {eqn:spin1}) is more involved. Assuming~(\ref{eqn:ECT1}) we have
\begin{equation}
d\left(e^\alpha\wedge e^\beta\right) + e^\gamma\wedge e^\beta\wedge\omega^\alpha{}_\gamma + e^\alpha\wedge e^\gamma\wedge \omega^\beta{}_\gamma=0\,.
\end{equation}
After expressing this equation in spinorial language, we decompose the connection one-forms according to (\ref{eqn:conndec}). Since $\alpha\mapsto AA'$ and $\beta\mapsto BB'$, we may now contract with $\epsilon_{A'B'}$. Due to symmetry reasons this cancels the connection terms involving $\bar\omega_{M'N'}$, and finally yields equation (\ref {eqn:spin1}). To see the converse (\ref {eqn:spin1})$\Rightarrow$(\ref{eqn:ECT1}), we define a covariant derivative
\begin{equation}\label{eqn:Ddef} 
De^{AA'}  = de^{AA'} - \omega^A{}_B\wedge e^{BA'} - \bar\omega^{A'}{}_{B'}\wedge e^{AB'}
\end{equation}
which allows us to rewrite (\ref {eqn:spin1}) as
\begin{equation}
2 De^{(M}{}_{Q'}\wedge e^{N)Q'} = 0\,. 
\end{equation}
This equation implies $De^{AA'}=0$. (One way to obtain this result is to expand $De^{AA'}$ in a basis of two-forms $e^{00'}\wedge e^{01'},\dots , e^{10'}\wedge e^{11'}$, observing that this quantity is Hermitian. One then substitutes the expansion into both the above equation and its complex conjugate, and compares coefficients.) Finally note that $De^{AA'}$ as defined in (\ref{eqn:Ddef}) is precisely the spinorial version of equation (\ref{eqn:ECT1}). 

This completes our proof of equivalence of the Lorentz tensor and the spinor formulation of Einstein--Cartan gravity. The latter will now become the basis for our review of Plebanski gravity.

\subsection{Plebanski gravity}
The key to proceed from the Einstein--Cartan action (\ref{eqn:ECTact2}) to Plebanski's formulation of gravity is the simple observation that the tetrads only appear in the combinations
\begin{equation}\label{eqn:defSi}
\Sigma^{AB}  = e^{AC'}\wedge e^B{}_{C'}\,,\qquad 
\bar\Sigma^{A'B'}  = e^{CA'}\wedge e_C{}^{B'}\,.
\end{equation}
These present a basis of two-forms in terms of which the tetrad-induced basis can be expressed as
\begin{equation}\label{eqn:eesi}
e^{AA'}\wedge e^{BB'} = -\frac{1}{2}\left(\epsilon^{A'B'}\Sigma^{AB}+\epsilon^{AB}\bar\Sigma^{A'B'}\right).
\end{equation}

Using the Infeld-van der Waerden map, one finds that the two-forms $\Sigma$ and $\bar \Sigma$ satisfy the following important geometric properties with respect to the spacetime Hodge star operator that acts as $\star (dx^a\wedge dx^b) = \sqrt{|g|}^{-1}\epsilon^{abpq}g_{pc}g_{qd} dx^c\wedge dx^d/2$ as an endomorphism of two forms: 
\begin{equation}
\star \Sigma^{AB} = -i \Sigma^{AB}\,,\qquad
\star\bar\Sigma^{A'B'} = +i \bar\Sigma^{A'B'}\,.
\end{equation}
With this sign convention, the $\Sigma^{AB}$ are called selfdual, and the $\bar\Sigma^{A'B'}$ anti-selfdual. From the two-form identity $\Sigma\wedge \star\bar\Sigma=\star\Sigma\wedge\bar\Sigma$   one finds the useful relation
\begin{equation}\label{eqn:ssbar}
\Sigma^{AB}\wedge \bar\Sigma^{C'D'} =0\,.
\end{equation}
Furthermore, the wedge product of two two-forms $\Sigma$ is proportional to the canonical volume form $d^4x\equiv dx^0\wedge dx^1\wedge dx^2\wedge dx^3$; one finds
\begin{equation}\label{eqn:sisi}
\Sigma^{AB}\wedge \Sigma_{CD} = 4i \,\delta^{(A}_C\delta^{B)}_D\sqrt{|g|} \,d^4x\,.
\end{equation}

In terms of the two-forms $\Sigma$ and $\bar \Sigma$, the Einstein--Cartan action integral~(\ref{eqn:ECTact2}) becomes
\begin{equation}
2i \int_M \Sigma^{AB}[e]\wedge R_{AB}[\omega] - \bar\Sigma^{A'B'}[e]\wedge\bar R_{A'B'}[\bar\omega] 
  + \frac{\lambda}{3} \Sigma^{AB}[e]\wedge\Sigma_{AB}[e]\,.
\end{equation}
Now Plebanski's idea was to regard the two-forms $\Sigma^{AB}$, the $SL(2,\mathbb{C})$ connection $\omega_{MN}$, and their complex conjugates, as new fundamental variables of the gravitational field. However, this does not yet result in a gravity theory equivalent to the original Einstein--Cartan theory because the above action (with $\Sigma$ no longer depending on $e$) and the resulting equations do not force the variables~$\Sigma$ and $\bar\Sigma$ to be simple wedge products of some suitable tetrad. Moreover, the equation of motion for~$\Sigma$ (in the simple case of cosmological constant $\lambda=0$) would imply zero curvature $R_{AB}[\omega]=0$, and so the theory could not yield an interesting gravity theory.

In order to repair these problems, it turns out to be sufficient to implement the constraint that ensures the existence of a tetrad $e^{AA'}$ so that the two-forms $\Sigma$ and $\bar\Sigma$ can be written as simple wedge products in the form of equations (\ref{eqn:defSi}). This constraint takes the form $\Sigma^{(AB}\wedge\Sigma^{CD)}=0$ as discussed in \cite{Capovilla:1991qb}. Accordingly, the action must be modified by a totally symmetric spinor-valued function $\Psi_{ABCD}$ that acts as a Lagrange multiplier,
\begin{eqnarray}\label{eqn:Pact}
S^\mathrm{P}[\Sigma,\bar\Sigma,\omega,\bar\omega,\Psi,\bar\Psi] & = & 2i \int_M \Sigma^{AB}\wedge R_{AB}[\omega] - \bar\Sigma^{A'B'}\wedge\bar R_{A'B'}[\bar\omega] \nonumber\\ 
&& \qquad - \frac{1}{2}\Psi_{ABCD}\Sigma^{AB}\wedge\Sigma^{CD} + \frac{1}{2}\bar\Psi_{A'B'C'D'}\bar\Sigma^{A'B'}\wedge\bar\Sigma^{C'D'} \nonumber\\
&& \qquad  + \frac{\lambda}{6} \Sigma^{AB}\wedge\Sigma_{AB} - \frac{\lambda}{6} \bar\Sigma^{A'B'}\wedge\bar\Sigma_{A'B'}\,.
\end{eqnarray}
This is the Plebanski action. The field equations obtained by variation with respect to $\omega$, $\Sigma$ and~$\Psi$ are
\begin{subequations}\label{eqn:023}
\begin{eqnarray}
d\Sigma^{AB}-2\omega^{(A}{}_C\wedge \Sigma^{B)C} &=&0\,,\label{eqn:Pleb1}\\
R_{AB}-\Psi_{ABCD}\Sigma^{CD}+\frac{\lambda}{3}\Sigma_{AB} &=& 0\,,\label{eqn:Pleb2}\\
\Sigma^{(AB}\wedge \Sigma^{CD)} &=&0\,.\label{eqn:Pleb3}
\end{eqnarray}
\end{subequations}
The equations for $\bar\omega$, $\bar\Sigma$ and $\bar\Psi$ turn out to be the complex conjugates. 

The vacuum Plebanski field equations (\ref{eqn:023}) are completely equivalent to the vacuum Einstein--Cartan field equations~(\ref{eqn:012}). We will recover this result in section~\ref{sec:matP} as a special case from our more general proof of equivalence for the non-vacuum case.

We remark that there are two possible ways of reconstructing the information about the spacetime metric from the two-forms $\Sigma^{AB}$ that solve the Plebanski equations. On the one hand, one can construct the tetrads in terms of which the two-forms can be written as in equation (\ref{eqn:defSi}), and then one calculates
\begin{equation}\label{eqn:ge}
g = e^{AA'} \otimes e_{AA'} = e^{AA'}{}_a e_{AA'\,b}\, dx^a\otimes dx^b\,.
\end{equation}
On the other hand, it is also possible to use the so-called Urbantke formula first discovered in~\cite{Urbantke:1984eb} to calculate the metric directly,
\begin{equation}\label{eqn:Urbantke}
\sqrt{|\mathrm{det}\,g|}\, g_{ab}dx^a\otimes dx^b = \frac{i}{24}\epsilon^{mnpq}\Sigma^{AB}{}_{am} \Sigma_B{}^C{}_{np} \Sigma_{CA\,qb} \,dx^a\otimes dx^b\,.
\end{equation}
The Urbantke formula will become important as a constructive tool in section~\ref{sec:matP} where we will present general consistent matter couplings to Plebanski theory both for tensor and for spinor fields. The formula is also important in recent work on geometric extensions of two-form gravity, as we will discuss briefly in the conclusion.

\section{Matter coupling to Einstein--Cartan gravity}\label{sec:ECmat}
So far we have discussed Einstein--Cartan and Plebanski gravity in vacuum. We now wish to complete these theories by coupling the common types of matter that either appear as tensor or as spinor fields. In this section we begin by discussing matter fields in Einstein--Cartan gravity. We do so in order to lay the foundations for our key results in the following section~\ref{sec:matP} where we develop new consistent matter couplings to Plebanski gravity. In particular we here introduce a systematic approach to extract the full content of the gravitational field equations. This will be essential later on when we will prove the full equivalence of non-vacuum Plebanski gravity with the new matter couplings to Einstein--Cartan gravity.

\subsection{Tensor fields}\label{sec:tenmat}
The matter actions for all common tensor fields, collectively denoted by $Q$, are of the generic form $S_\mathrm{m}[g,Q]$, and do not depend on covariant derivatives related to the metric $g$. Using formula~(\ref{eqn:ge}) immediately provides a minimal coupling of fields $Q$ to Einstein--Cartan gravity; the action becomes 
\begin{equation}
S_\mathrm{m}^\mathrm{EC}[e,Q] = S_\mathrm{m}[g(e),Q]\,.
\end{equation}
The full gravitational field equations of Einstein--Cartan gravity are obtained by variation of the total action $S^\mathrm{EC}+S_\mathrm{m}^\mathrm{EC}$, see (\ref{eqn:ECTact2}), with respect to the tetrads $e$ and the $SL(2,\mathbb{C})$-connections~$\omega$ and~$\bar\omega$. Matter source terms arise from
\begin{equation}\label{eqn:var1}
\delta_e S_\mathrm{m}^\mathrm{EC} = \int_M d^4x \frac{\delta S_\mathrm{m}}{\delta g_{ab}}\frac{\delta g_{ab}}{\delta e^{AA'}_p}\delta e^{AA'}_p = \int_M d^4x\sqrt{|g|}\,\frac{1}{2} T^{ab} \frac{\delta g_{ab}}{\delta e^{AA'}_p} e^{CC'}_p \delta e^{AA'}{}_{CC'}\,,
\end{equation}
where we used the definition
\begin{equation}\label{eqn:em}
T^{ab}=\frac{2}{\sqrt{|g|}}\frac{\delta S_\mathrm{m}}{\delta g_{ab}} 
\end{equation}
for the matter energy momentum tensor, and reexpressed the spacetime components $\delta e^{AA'}_p$ of the variation $\delta e^{AA'}$ in terms of components $\delta e^{AA'}{}_{CC'}$ with respect to the tetrad basis. Combining this result with the vacuum field equations (\ref{eqn:012}) results in
\begin{subequations}
\begin{eqnarray}
0 &=& d\left(e^{MA'}\wedge e^N{}_{A'}\right) - 2 \omega^{(M}{}_A\wedge e^{N)A'}\wedge e^A{}_{A'}\,,\label{eqn:ecm1}\\
0 &=& e^{CC'}\wedge e^{AA'} \!\wedge \!\left( R_{MA}[\omega]\epsilon_{M'A'} - \bar R_{M'A'}[\bar\omega]\epsilon_{MA}\right)\nonumber\\ 
&&{}- \frac{2\lambda}{3} e^{CC'}\wedge e^A{}_{M'}\wedge e_M{}^{A'}\wedge e_{AA'}
+\frac{i}{4}\sqrt{|g|}\,d^4x\,T^{ab}e_{aMM'}e^{CC'}_b\,.\label{eqn:ecm2}
\end{eqnarray}
\end{subequations}

We will now systematically extract the content of the second field equation by calculating its four-form components. For this purpose we insert the two-form basis decomposition
\begin{equation}\label{eqn:rdec}
R_{AB}=\left(\psi_{ABCD}+\Lambda\epsilon_{(A|C|}\epsilon_{B)D}\right)\Sigma^{CD}+\Phi_{ABC'D'}\bar\Sigma^{C'D'}\,.
\end{equation}
The component functions $\psi_{ABCD}$ are totally symmetric in their spinor indices, and correspond to the Weyl tensor part in the spinor decomposition of the Riemann curvature. Moreover, the Hermitian components $\Phi_{ABC'D'}=\bar\Phi_{ABC'D'}$ encode the tracefree Ricci tensor, while the real function $\Lambda=\bar\Lambda$ is determined by the Ricci scalar~\cite{Plebanski:1975wn}. We then express all wedge products of two tetrads in terms of $\Sigma$ and $\bar\Sigma$ via formula (\ref{eqn:eesi}). Due to (\ref{eqn:ssbar}) only terms of the form $\Sigma\wedge\Sigma$ and $\bar\Sigma\wedge \bar\Sigma$ remain; these can be simplified using (\ref{eqn:sisi}). Following these steps, and using the facts that, by construction,~$\Phi$ is Hermitian and $\Lambda$ is real, 
one obtains the following component equation equivalent to (\ref{eqn:ecm2}),
\begin{equation}\label{eqn:ecm}
2\Phi_{MNM'N'} + \left(\lambda+3\Lambda\right)\epsilon_{MN}\epsilon_{M'N'} = \frac{1}{8}T_{MNM'N'}\,,
\end{equation}
where we denote the components of the matter energy momentum tensor in the tetrad basis by
\begin{equation}\label{eqn:Tt}
T^{MNM'N'}=T^{ab}e^{MM'}_ae^{NN'}_b\,.
\end{equation}
We finally decompose equation (\ref{eqn:ecm}) into components of independent symmetry. To do so we make use of the fact that expressions $E_{AB}=-E_{BA}$ that are antisymmetric in two spinor indices must be proportional to the symplectic form $\epsilon_{AB}$; more precisely, one finds that $2E_{AB}=-E^C{}_{C}\epsilon_{AB}$. Eventually, the symmetry decomposition yields
\begin{subequations}\label{eqn:033}
\begin{eqnarray}
0 &=& \frac{1}{16} T_{(MN)(M'N')} -\Phi_{MNM'N'}\,,\label{eqn:ecms1}\\
0 &=& \frac{1}{32} T^{CC'}{}_{CC'}-\lambda-3\Lambda\,.\label{eqn:ecms2}
\end{eqnarray}
\end{subequations}
This determines the curvature two-form
\begin{equation}
R_{AB} = \psi_{ABCD}\Sigma^{CD} -\frac{1}{96}\big(32\lambda-T^{CC'}{}_{CC'}\big)\Sigma_{AB}+\frac{1}{16}T_{(AB)(C'D')}\bar\Sigma^{C'D'}
\end{equation}
in terms of the theory's matter content and the cosmological constant $\lambda$. The Weyl contribution $\psi_{ABCD}$ is not restricted by the equations of motion, as is the case in general relativity.

The two equations (\ref{eqn:033}) together with the vanishing torsion condition (\ref{eqn:ecm1}) concisely summarize the full dynamical content of Einstein--Cartan gravity coupled to tensor field matter.

\subsection{Spinor fields}\label{sec:ECspin}
In contrast to the situation for tensor fields, the action of spinor fields $\zeta$ in Einstein--Cartan gravity also depends on the connection one-forms that directly couple to the spinor fields via the Lorentz covariant derivative. The action has the generic structure 
\begin{equation}
S_\zeta^\mathrm{EC}[e,\omega,\bar\omega,\zeta]\,.
\end{equation}
Consequently, one now obtains source terms from the variations with respect to the tetrad $e$ and the connections $\omega$ and $\bar\omega$. Below we only display the arguments related to $\omega$; those related to $\bar\omega$ are similar. From
\begin{equation}
\delta_\omega S_\zeta^\mathrm{EC} = \int_M d^4x \frac{\delta S_\zeta^\mathrm{EC}}{\delta \omega_{ABp}}\delta \omega_{ABp} = \int_M d^4x \frac{\delta S_\zeta^\mathrm{EC}}{\delta \omega_{ABp}}e_p^{CC'}\delta \omega_{ABCC'}
\end{equation}
we find the torsion condition
\begin{equation}\label{eqn:modtor}
0 = \left[d\left(e^{MA'}\wedge e^N{}_{A'}\right) - 2 \omega^{(M}{}_A\wedge e^{N)A'}\wedge e^A{}_{A'}\right]\wedge e^{CC'} + \frac{i}{2}d^4x\frac{\delta S_\zeta^\mathrm{EC}}{\delta \omega_{ABp}}e_p^{CC'}\,.
\end{equation}

The variation
\begin{equation}
\delta_e S_\zeta^\mathrm{EC} = \int_M d^4x \frac{\delta S_\zeta^\mathrm{EC}}{\delta e^{AA'}_p}\delta e^{AA'}_p = \int_M d^4x \frac{\delta S_\zeta^\mathrm{EC}}{\delta e^{AA'}_p}e^{CC'}_p \delta e^{AA'}{}_{CC'}
\end{equation}
is structurally very similar to the corresponding variation (\ref{eqn:var1}) for tensor field matter; the only  difference is that the term $\sqrt{|g|}T^{ab}\,\delta g_{ab}/\delta e^{AA'}_p$ there is replaced by $2 \,\delta S_\zeta^\mathrm{EC}/\delta e^{AA'}_p$ here. Hence the equation of motion is (\ref{eqn:ecm2}) with the same replacement. In the same way as before we may then systematically extract the four-form components of this equation, which yields
\begin{equation}
2\Phi_{MNM'N'} + \left(\lambda+3\Lambda\right)\epsilon_{MN}\epsilon_{M'N'} = \frac{1}{8}\tilde T_{MNM'N'}
\end{equation}
where now
\begin{equation}\label{eqn:Ttilde}
\tilde T_{MNM'N'} = \frac{1}{\sqrt{|g|}}\frac{\delta S_\zeta^\mathrm{EC}}{\delta e^{MM'}_p} e_{NN'p}\,.
\end{equation}
The decomposition of this equation into components of independent symmetry finally gives 
\begin{subequations}\label{eqn:041}
\begin{eqnarray}
0 &=& \frac{1}{16} \tilde T_{(MN)(M'N')} -\Phi_{MNM'N'}\,,\label{eqn:11}\\
0 &=& \frac{1}{32} \tilde T^{CC'}{}_{CC'}-\lambda-3\Lambda\,,\label{eqn:12}\\
0 &=& \tilde T_{(MN)}{}^{C'}{}_{C'}\,.\label{eqn:13}
\end{eqnarray}
\end{subequations}
Comparison to (\ref{eqn:033}) shows that the third condition does not occur for tensor field matter. Here it arises because $\tilde T_{MNM'N'}$ is not by definition symmetric under the interchange of index pairs $MM'$ and $NN'$.

Collectively denoting by $\zeta$ both tensor and spinor field matter, we thus conclude that the full dynamical content of Einstein--Cartan gravity is captured by the torsion condition (\ref{eqn:modtor}), a similar equation derived by variation with respect to $\bar\omega$, and the relations (\ref{eqn:041}) between curvature and energy momentum.

\section{Matter coupling to Plebanski gravity}\label{sec:matP}
This section contains the central results of this article. Most importantly, we will present a new scheme for the minimal coupling of tensor field matter to Plebanski gravity. We will prove the equivalence of the resulting theory to Einstein--Cartan gravity with matter by comparison to the results of the previous section. We will also reconsider the proposals for scalar and Yang--Mills couplings from~\cite{Capovilla:1991qb}; their full consistency will be established for the first time. Finally, we will show how to extend our new minimal coupling scheme so that also spinor fields can be included consistently in Plebanski gravity. 

\subsection{New minimal coupling scheme for tensor field matter}\label{sec:minc}
As discussed in section~\ref{sec:ECmat}, the generic matter action for tensor fields $Q$ only depends on the metric $g$ and takes the form $S_\mathrm{m}[g,Q]$. The Urbantke formula (\ref{eqn:Urbantke}) now provides an expression for the spacetime metric in terms of two-forms $\Sigma^{AB}$. We can also use the complex conjugate of this formula to obtain an alternative expression for $g$ in terms of the $\bar\Sigma^{A'B'}$. 

Based on this observation we propose the following natural minimal coupling for tensor field matter to Plebanski gravity: 
\begin{equation}\label{eqn:minc3}
S_\mathrm{m}^\mathrm{P}[\Sigma,\bar\Sigma,Q] = \frac{1}{2}S_\mathrm{m}[g(\Sigma),Q]+\frac{1}{2}S_\mathrm{m}[g(\bar\Sigma),Q]\,.
\end{equation}

This action is chosen to be symmetric in the $\Sigma$ and $\bar\Sigma$, so that both corresponding variation equations will acquire source terms. We write
\begin{eqnarray}
\delta_\Sigma S_\mathrm{m}^\mathrm{P} &=& \int_M d^4x \frac{1}{2} \frac{\delta S_\mathrm{m}}{\delta g_{ab}} \frac{\delta g_{ab}}{\delta \Sigma^{AB}_{pq}}\delta \Sigma^{AB}_{pq}\nonumber\\
&=& \int_M d^4x\sqrt{|g|} \,\frac{1}{4} T^{ab} \frac{\delta g_{ab}}{\delta \Sigma^{AB}_{pq}}\!\left(\Sigma^{PQ}_{pq}\delta\Sigma^{AB}{}_{PQ}+\bar\Sigma^{P'Q'}_{pq}\delta\Sigma^{AB}{}_{P'Q'}\right)
\end{eqnarray}
using our definition (\ref{eqn:em}) of the metric energy momentum tensor and expressing the variation $\delta \Sigma^{AB}_{pq}$ in terms of components with respect to the two-form basis given by $\Sigma,\bar\Sigma$. All these components appear because the variation $\delta \Sigma^{AB}_{pq}$ can contain both selfdual and antiselfdual contributions.

With this result we obtain the full set of gravitational field equations from the variation of the total action $S^\mathrm{P}+S_\mathrm{m}^\mathrm{P}$, see (\ref{eqn:Pact}), with respect to $\omega$, $\Sigma$ and the Lagrange multiplier $\Psi$:
\begin{subequations}
\begin{eqnarray}
d\Sigma^{AB}-2\omega^{(A}{}_C\wedge \Sigma^{B)C} &=&0\,,\label{eqn:mP1}\\
\Sigma^{PQ}\wedge\Big(R_{AB}-\Psi_{ABCD}\Sigma^{CD}+\frac{\lambda}{3}\Sigma_{AB}\Big) &=& \frac{i}{8} \sqrt{|g|}d^4x \,T^{ab} \frac{\delta g_{ab}}{\delta \Sigma^{AB}_{pq}} \Sigma^{PQ}_{pq}\,,\label{eqn:mP2}\\
\bar\Sigma^{P'Q'}\wedge\Big(R_{AB}-\Psi_{ABCD}\Sigma^{CD}+\frac{\lambda}{3}\Sigma_{AB}\Big) &=& \frac{i}{8} \sqrt{|g|}d^4x \,T^{ab} \frac{\delta g_{ab}}{\delta \Sigma^{AB}_{pq}} \bar\Sigma^{P'Q'}_{pq}\!,\label{eqn:mP3}\\
\Sigma^{(AB}\wedge \Sigma^{CD)} &=&0\,.\label{eqn:mP4}
 \end{eqnarray}
 \end{subequations}
The equations for $\bar\omega$ and $\bar\Psi$ are once again found to be the complex conjugates; this also holds for the equation for  $\bar\Sigma$ assuming real energy momentum $T^{ab}$. We will further comment on this point in section~\ref{sec:Pspin}. Note that the energy momentum tensor and the volume density depend on $g(\Sigma)$.

The final missing ingredient in the Plebanski field equations above is the term $\delta g_{ab}/\delta \Sigma^{AB}_{pq}$ that can be calculated from the Urbantke formula (\ref{eqn:Urbantke}) and from (\ref{eqn:sisi}). We also employ the selfduality condition for $\Sigma$ which reads
\begin{equation}\label{eqn:sdc}
\frac{1}{2\sqrt{|g|}}\epsilon^{abcd}\Sigma_{ABcd} = -i\,\Sigma_{AB}^{ab}
\end{equation}
in components. After some amount of simplification we obtain
\begin{eqnarray}\label{eqn:gsi}
\frac{\delta g_{ab}}{\delta \Sigma^{AB}_{pq}} &=& \underbrace{\frac{1}{\sqrt{|g|}}\frac{\delta}{\delta \Sigma^{AB}_{pq}}\left(\sqrt{|g|}g_{ab}\right)}_1 - \underbrace{\frac{g_{ab}}{\sqrt{|g|}} \frac{\delta}{\delta \Sigma^{AB}_{pq}} \sqrt{|g|}}_2\\
&=& \bigg[\underbrace{\frac{1}{12} \delta^{[p}_a \Sigma^{q]sR}{}_{(A}\Sigma_{B)Rsb} +\frac{i}{48}\sqrt{|g|}^{-1}\epsilon^{pqmn}\Sigma_{am}{}^R{}_{(A}\Sigma_{B)Rbn}}_1-\underbrace{\frac{1}{24}g_{ab}\Sigma^{pq}_{AB}}_2\bigg]+(a\leftrightarrow b)\,.\nonumber
\end{eqnarray}

Based on these results we will now prove the equivalence of Plebanski gravity with tensor field matter  minimally coupled as proposed in (\ref{eqn:minc3}) to Einstein--Cartan gravity.

\subsection{Proof of equivalence to non-vacuum Einstein--Cartan theory}\label{sec:peqec}
To prove the full equivalence of Einstein--Cartan and Plebanski gravity with tensor field matter, we recall that the constraint (\ref{eqn:mP4}) is equivalent to the statement that the two-forms $\Sigma$  are simple, i.e., induced by some tetrad of one-forms $e^{AA'}$ as $\Sigma^{AB}=e^{AC'}\wedge e^B{}_{C'}$.  In consequence, equation (\ref{eqn:mP1}) is immediately equivalent to the torsion-free condition (\ref{eqn:ecm1}) in Einstein--Cartan-gravity. It remains to be shown that the curvature equations (\ref{eqn:mP2}) and (\ref{eqn:mP3}) of Plebanski gravity are equivalent to the curvature equation in Einstein--Cartan gravity from which we extracted the dynamical content~(\ref{eqn:033}).

We proceed by simplifying the source terms in the Plebanski curvature equations under the assumption of simple two-forms $\Sigma$ and $\bar\Sigma$. For this purpose we employ the selfduality relation (\ref{eqn:sdc}), the component expressions 
\begin{equation}
\Sigma^{AB}_{pq} = 2 e^{AC'}_{[p} e_{q]}^B{}_{C'}\,,\qquad \bar\Sigma^{A'B'}_{pq} = 2e^{CA'}_{[p}e_{q]C}{}^{B'}\,, 
\end{equation}
and the identity
\begin{equation}
e^{AA'}_s e^{sBB'} = \epsilon^{AB}\epsilon^{BB'}\,.
\end{equation}
A lengthy expansion of all terms then results in 
\begin{subequations}\label{eqn:049}
\begin{eqnarray}
\frac{1}{\sqrt{|g|}}\frac{\delta}{\delta \Sigma^{AB}_{pq}}\left(\sqrt{|g|}g_{ab}\right)\Sigma^{PQ}_{pq} &=& \frac{1}{6}\delta^P_{(A}\delta^Q_{B)}g_{ab}+\frac{5}{3}e^{S'(P}_{(a}e^{}_{b)S'(A}\delta^{Q)}_{B)}\,,\label{eqn:3}\\
\frac{1}{\sqrt{|g|}}\frac{\delta}{\delta \Sigma^{AB}_{pq}}\left(\sqrt{|g|}g_{ab}\right)\bar\Sigma^{P'Q'}_{pq} &=& - 2\, e_{A}{}^{(P'}{}_{\!(a}e_{b)B}{}^{Q')}\,.\label{eqn:4}
\end{eqnarray}
\end{subequations}
Substituting these expressions into equation (\ref{eqn:gsi}) and using the definition (\ref{eqn:Tt}), we find that the curvature equations (\ref{eqn:mP2}) and (\ref{eqn:mP3}) reduce to
\begin{eqnarray}\label{eqn:1}
&&\Sigma^{PQ}\wedge\Big(R_{AB}-\Psi_{ABCD}\Sigma^{CD}+\frac{\lambda}{3}\Sigma_{AB}\Big)\\
\!\!\!\!{}&=& \frac{i}{48}\left(\sqrt{|g|}d^4x\, \delta^P_{(A}\delta^Q_{B)}+i \Sigma^{PQ}\wedge \Sigma_{AB}\right)T^{RS'}{}_{RS'} +\frac{5i}{24}\sqrt{|g|}d^4x\, \delta^{(P}_{(A}T^{Q)S'}{}_{B)S'}\nonumber
\end{eqnarray}
and
\begin{equation}\label{eqn:2}
\bar\Sigma^{P'Q'}\wedge\Big(R_{AB}-\Psi_{ABCD}\Sigma^{CD}+\frac{\lambda}{3}\Sigma_{AB}\Big)=-\frac{i}{4}\sqrt{|g|}d^4x\,T_{(AB)}{}^{(P'Q')}
\end{equation}

From these two equations we may now systematically extract the four-form components in precisely the same way as presented in section \ref{sec:tenmat}. The first equation (\ref{eqn:1}) is equivalent to the component expression
\begin{equation}
0 = -32 (\psi_{ABCD}-\Psi_{ABCD}) +\epsilon_{(A|C|}\epsilon_{B)D}\Big(-\frac{8}{3}(\lambda+3\Lambda)+\frac{1}{12}T^{RS'}{}_{RS'}\Big)\,,
\end{equation}
which, after symmetry decomposition, tells us to identify the Lagrange multiplier $\Psi$ with the Weyl part $\psi$ in the curvature decomposition (\ref{eqn:rdec}), but does not at all constrain the Weyl part. Otherwise the expression above is equivalent to equation (\ref{eqn:ecms2}). The second equation (\ref{eqn:2}) is equivalent to the following component expression which is in turn equivalent to (\ref{eqn:ecms1}):
\begin{equation}
0= 8\Phi_{ABC'D'}-\frac{1}{2}T_{(AB)(C'D')}\,.
\end{equation}

This completes the proof of equivalence of Einstein--Cartan gravity and Plebanski gravity with tensor field matter minimally coupled as proposed in the preceding section \ref{sec:minc}. As an important special case one immediately recovers the equivalence of both theories in vacuum.

\subsection{Consistency of known couplings}
With our new technology for matter coupling to Plebanski gravity firmly in place, we are now in the position to investigate the consistency of the actions for scalar and Yang--Mills fields previously suggested in the literature~\cite{Capovilla:1991qb}. 

We first consider the massless scalar field action 
\begin{equation}
S_\phi^\mathrm{P}[\Sigma,\bar\Sigma,\phi,\pi]=\frac{1}{2}\int_M d^4x\sqrt{|g(\Sigma)|}\Big(\pi^a\partial_a\phi-\frac{1}{2}g(\Sigma)_{ab}\pi^a\pi^b\Big) + \left(\Sigma\mapsto\bar\Sigma\right)
\end{equation}
in symmetric form with respect to $\Sigma$ and $\bar\Sigma$. The presence of the auxiliary field $\pi^a$, that is fixed by the equations of motion to be $g(\Sigma)_{ab}\pi^b=\partial_a\phi$ which in turn yields the Klein--Gordon equation for $\phi$, is motivated in~\cite{Capovilla:1991qb} by the desire to keep the action polynomial in the two-forms $\Sigma,\bar\Sigma$. This is achieved through the Urbantke formula (\ref{eqn:Urbantke}) and relation (\ref{eqn:sisi}). Observe that the action in the above form is precisely of the minimal coupling type we proposed in (\ref{eqn:minc3}). Hence the results of this article for the first time prove the consistency of this scalar field action within Plebanski gravity.

Second, we consider the Yang--Mills action suggested in~\cite{Capovilla:1991qb},
\begin{eqnarray}\label{eqn:YMC}
S_\mathrm{YM}^\mathrm{P}[\Sigma,\bar\Sigma,A,\phi,\bar\phi] &=& i\int_M \mathrm{Tr}\Big(F\wedge \Sigma^{AB}\phi_{AB} -\frac{1}{2}\phi_{AB}\phi_{CD}\Sigma^{AB}\wedge \Sigma^{CD}\Big.\nonumber\\
&& \Big.\qquad\qquad- F\wedge \bar\Sigma^{A'B'}\bar\phi_{A'B'}+\frac{1}{2}\bar\phi_{A'B'}\bar\phi_{C'D'}\bar\Sigma^{A'B'}\wedge \bar\Sigma^{C'D'}\Big),
\end{eqnarray}
that depends on a Lie algebra valued gauge potential $A$ with real field strength $F=dA+A\wedge A$ and an auxiliary spinor valued symmetric function $\phi_{AB}$. The trace is that of the Lie algebra elements in the adjoint representation, i.e., the Lie algebra inner product given by the Killing form. We write the action above in symmetrized form with respect to $\Sigma,\bar\Sigma$ in order to make it real. As argued in~\cite{Capovilla:1991qb}, the auxiliary field can be eliminated from the action so that the equations of motion take the standard Yang--Mills form. But consistency of the gravitational coupling was not shown; neither is it immediate from our results above, since the action is not of minimal coupling type~(\ref{eqn:minc3}). Nevertheless we will now demonstrate that the action above is indeed consistent.

Variation of $S_\mathrm{YM}^\mathrm{P}$ with respect to the auxiliary fields $\phi_{AB}$ and $\bar\phi_{A'B'}$, by employing the two-form basis decomposition
\begin{equation}\label{eqn:Fbas}
F=F_{AB}\Sigma^{AB}+\bar F_{A'B'}\bar\Sigma^{A'B'}\,,
\end{equation}
determines $\phi_{AB}=F_{AB}$ and $\bar\phi_{A'B'}=\bar F_{A'B'}$. The gravitational field equations in this case are obtained from the total action $S^\mathrm{P}+S_\mathrm{YM}^\mathrm{P}$, see (\ref{eqn:Pact}). We only need to consider the variation with respect to $\Sigma$ to see the modifications of our discussion in the previous sections \ref{sec:minc} and \ref{sec:peqec}. With the result above for the auxiliary fields we obtain 
\begin{subequations}
\begin{eqnarray}
\Sigma^{PQ}\wedge\Big(R_{AB}-\Psi_{ABCD}\Sigma^{CD}+\frac{\lambda}{3}\Sigma_{AB}\Big) &=& 0\,,\\
\bar\Sigma^{P'Q'}\wedge\Big(R_{AB}-\Psi_{ABCD}\Sigma^{CD}+\frac{\lambda}{3}\Sigma_{AB}\Big) &=& -\frac{1}{2}F_{AB}\bar F_{C'D'}\bar\Sigma^{P'Q'}\wedge\bar\Sigma^{C'D'}\,,
\end{eqnarray}
\end{subequations}
which should be compared to equations (\ref{eqn:mP2}) and (\ref{eqn:mP3}). The systematic extraction of four-form components and a split of the resulting equation into terms of independent symmetry makes these equivalent to the equation $\Psi_{ABCD}=\psi_{ABCD}$ that fixes the Lagrange multiplier $\Psi$ (but does not constrain the Weyl curvature $\psi$) and to
\begin{equation}
\Phi_{ABC'D'}+\frac{1}{2}F_{AB}\bar F_{C'D'}=0\,,\qquad \lambda+3\Lambda=0\,.
\end{equation}

These equations must be compared to (\ref{eqn:ecms1}) and (\ref{eqn:ecms2}). We conclude that the Yang--Mills coupling $S_\mathrm{YM}^\mathrm{P}$ is consistent by equivalence to Einstein--Cartan gravity if and only if the Yang--Mills energy momentum has the components
\begin{equation}
T_{(MN)(M'N')} = -8F_{MN}\bar F_{M'N'}\,,\qquad T^{PP'}{}_{PP'}=0\,.
\end{equation}
That this is indeed the case can be seen by mapping the well-known expression for the metric Yang--Mills energy momentum tensor
\begin{equation}
T^{ab} = \mathrm{Tr}\Big(F^{ap}F^b{}_p-\frac{1}{4} g^{ab}F^{pq}F_{pq} \Big),
\end{equation}
to $T^{MNM'N'}$ according to (\ref{eqn:Tt}), by then using the basis decomposition (\ref{eqn:Fbas}), and finally by expanding all occurring two-forms in a tetrad frame. 

Hence the Yang--Mills action (\ref{eqn:YMC}) consistently couples to Plebanski gravity and provides an alternative rewriting of the Yang--Mills field minimally coupled according to (\ref{eqn:minc3}).  However, our  general proposal has the clear advantage that it does not need to be checked case by case.

\subsection{Coupling of spinor fields and equivalence}\label{sec:Pspin}
We could show in the preceding section that the couplings for scalar and Yang--Mills fields suggested in~\cite{Capovilla:1991qb} at the level of the respective matter actions are also consistent gravitational couplings in Plebanski gravity. The same article also presented an `artificial' candidate for a spinor coupling, but we will not analyze the consistency of this ansatz here. Instead, we will show in this section how to extend our newly proposed minimal coupling scheme consistently to spinor field matter.

The basic observation we need is that tetrad frames can be determined from a given spacetime metric up to a local Lorentz transformation. The spacetime metric in turn can be determined via the Urbantke formula from either set of two-forms $\Sigma^{AB}$ or $\bar\Sigma^{A'B'}$. Starting from the Einstein--Cartan spinor action $S_\zeta^\mathrm{EC}[e,\omega,\bar\omega,\zeta]$, see section \ref{sec:ECspin}, we thus can construct a minimal coupling of spinors to Plebanski gravity as
\begin{equation}\label{eqn:minc2}
S_\zeta^\mathrm{P}[\Sigma,\bar\Sigma,\omega,\bar\omega,\zeta] = \frac{1}{2}S_\zeta^\mathrm{EC}[e(g(\Sigma)),\omega,\bar\omega,\zeta] +\frac{1}{2}S_\zeta^\mathrm{EC}[e(g(\bar\Sigma)),\omega,\bar\omega,\zeta]\,.  
\end{equation}

From this form of the action we may determine a number of useful relations between the variations with respect to $\Sigma$ and $\bar\Sigma$ in case they are induced by a tetrad frame. These will be needed below in order to prove the equivalence of Einstein--Cartan and Plebanski gravity including spinor fields. First, denoting by $T^{ab}$ the formal real energy momentum tensor of $S_\zeta^\mathrm{EC}[e(g),\omega,\bar\omega,\zeta]$, one finds that
\begin{equation}
T^\Sigma{}^{pq}_{AB} = \frac{1}{\sqrt{|g|}}\frac{\delta S_\zeta^\mathrm{P}}{\delta \Sigma^{AB}_{pq}}=\frac{1}{4} T^{ab} \frac{\delta g_{ab}}{\delta \Sigma^{AB}_{pq}}\,,
\end{equation}
is the conjugate of 
\begin{equation}
T^{\bar\Sigma}{}^{pq}_{A'B'} = \frac{1}{\sqrt{|g|}}\frac{\delta S_\zeta^\mathrm{P}}{\delta \bar\Sigma^{A'B'}_{pq}}=\frac{1}{4} T^{ab} \frac{\delta g_{ab}}{\delta \bar\Sigma^{A'B'}_{pq}}\,.
\end{equation}
In other words, $\overline{T^\Sigma}=T^{\bar\Sigma}$. To see this, one verifies that $\delta g/\delta \bar\Sigma$ is the conjugate of $\delta g/\delta\Sigma$ by a similar calculation as led to (\ref{eqn:gsi}). Second, we may consider the identities
\begin{subequations}
\begin{eqnarray}
\frac{\delta g_{ab}}{\delta \Sigma^{AB}_{pq}}\Sigma^{AB}_{pq} &=& g_{ab}\,,\\
\frac{\delta g_{ab}}{\delta \Sigma^{AB}_{pq}}\bar\Sigma^{P'Q'}_{pq} &=& -2\,e_A{}^{(P'}{}_{(a}e_{b)B}{}^{Q')}\,,
\end{eqnarray}
\end{subequations}
which follow from combining our previous results in (\ref{eqn:gsi}) with (\ref{eqn:049}). These imply that $T^\Sigma{}^{pq}_{AB}\Sigma^{AB}_{pq}$ is real and $T^\Sigma{}^{pq}_{AB}\bar\Sigma^{P'Q'}_{pq}$ is Hermitian. To summarize, we have that
\begin{equation}\label{eqn:5}
T^\Sigma{}^{pq}_{AB}\Sigma^{AB}_{pq}=T^{\bar\Sigma}{}^{pq}_{A'B'}\bar\Sigma^{A'B'}_{pq}\,,\qquad
T^\Sigma{}^{pq}_{PQ}\bar\Sigma^{A'B'}_{pq}=T^{\bar\Sigma}{}^{pqA'B'}\Sigma_{PQpq}\,.
\end{equation}

The gravitational field equations are derived by variation of the total action $S^\mathrm{P} +S_\zeta^\mathrm{P} $, see (\ref{eqn:Pact}) and (\ref{eqn:minc2}), with respect to $\omega$, $\Sigma$ and the Lagrange multiplier $\Psi$. We obtain source terms from 
\begin{subequations}
\begin{eqnarray}
\delta_\omega S_\zeta^\mathrm{P} &=& \int_M d^4x \frac{\delta S_\zeta^\mathrm{P}}{\delta \omega_{ABp}} e^{CC'}_p \delta \omega_{ABCC'}\,,\\
\delta_\Sigma S_\zeta^\mathrm{P} &=& \int_M d^4x \frac{\delta S_\zeta^\mathrm{P}}{\delta \Sigma^{AB}_{pq}} \!\left(\Sigma^{PQ}_{pq}\delta\Sigma^{AB}{}_{PQ}+\bar\Sigma^{P'Q'}_{pq}\delta\Sigma^{AB}{}_{P'Q'}\right).
\end{eqnarray}
\end{subequations}
Hence the resulting equations read
\begin{subequations}
 \begin{eqnarray}
\Big(d\Sigma^{AB}-2\omega^{(A}{}_P\wedge \Sigma^{B)P}\Big)\wedge e^{CC'} &=& -\frac{i}{2}d^4x\,\frac{\delta S_\zeta^\mathrm{P}}{\delta \omega_{ABp}} e^{CC'}_p\,,\label{eqn:mPs1}\\
\Sigma^{PQ}\wedge\Big(R_{AB}-\Psi_{ABCD}\Sigma^{CD}+\frac{\lambda}{3}\Sigma_{AB}\Big) &=& \frac{i}{2} \sqrt{|g|}d^4x \,T^\Sigma{}^{pq}_{AB} \Sigma^{PQ}_{pq}\,,\label{eqn:mPs2}\\
\!\!\!\!\!\!\!\!\bar\Sigma^{P'Q'}\wedge\Big(R_{AB}-\Psi_{ABCD}\Sigma^{CD}+\frac{\lambda}{3}\Sigma_{AB}\Big) &=& \frac{i}{2} \sqrt{|g|}d^4x \,T^\Sigma{}^{pq}_{AB}\bar\Sigma^{P'Q'}_{pq}\!,\label{eqn:mPs3}\\
\Sigma^{(AB}\wedge \Sigma^{CD)} &=&0\,.\label{eqn:mPs4}
 \end{eqnarray}
 \end{subequations}
As usual, the torsion condition is modified by the appearance of spinor fields. The equation for $\bar\omega$ is a similar condition; we do not display it here since it does not affect our proof of equivalence. The equations obtained by variation with respect to $\bar\Psi$ is the conjugate of the simplicity constraint above. Moreover, using identities (\ref{eqn:5}), one finds that the equations obtained by variation with respect to $\bar\Sigma$ are the conjugates of those obtained for $\Sigma$. This fact was also used in our calculation for tensor fields.

Note that one may define two forms $E^\Sigma$ from the energy momentum components $T^\Sigma$ according to $E^\Sigma_{ABab}dx^a\wedge dx^b=\sqrt{|g|}\epsilon_{abpq}T^\Sigma{}^{pq}_{AB}dx^a\wedge dx^b$; and similarly for $E^{\bar\Sigma}$. These correspond to the energy momentum two-forms discussed in~\cite{Krasnov:2008ui}. In terms of these the variation of the matter action becomes
\begin{equation}
\delta_\Sigma S_\zeta^\mathrm{P}  = \int_M E^\Sigma_{AB}\wedge \delta\Sigma^{AB}\,.
\end{equation}
Equations (\ref{eqn:mPs2}) and (\ref{eqn:mPs3}) can then be rewritten as
\begin{equation}
R_{AB}-\Psi_{ABCD}\Sigma^{CD}+\frac{\lambda}{3}\Sigma_{AB} = -E^\Sigma_{AB}\,,
\end{equation}
but this form is less suitable for our argument below.

Following the systematic procedure presented in section \ref{sec:tenmat} we extract the equivalent four-form component equations from (\ref{eqn:mPs2}) and (\ref{eqn:mPs3}). The decomposition into equations of independent symmetry yields:
\begin{subequations}
\begin{eqnarray}
0 & = & \psi_{ABCD} - \Psi_{ABCD} - \frac{1}{8}T^\Sigma{}^{pq}_{(AB}\Sigma_{CD)pq}\,,\label{eqn:6}\\
0 & = & \Phi_{ABC'D'} +\frac{1}{8} T^\Sigma{}^{pq}_{AB} \bar\Sigma_{C'D'pq}\,,\label{eqn:7}\\
0 & = & \lambda+3\Lambda -\frac{1}{8}T^\Sigma{}^{pq}_{PQ}\Sigma^{PQ}_{pq}\,,\label{eqn:8}\\
0 & = & T^\Sigma{}^{pq}_{P(A}\Sigma^P{}_{B)pq}\,.\label{eqn:9}
\end{eqnarray}
\end{subequations}
The full dynamical content of Plebanski gravity with spinor field matter (and tensor field matter)~$\zeta$ is now nicely summarized by these equations together with the torsion condition (\ref{eqn:mPs1}) and the simplicity constraint (\ref{eqn:mPs4}).

In order to prove the equivalence of this formulation to Einstein--Cartan gravity we need to translate expressions involving $T^\Sigma$ into expressions in terms of $\tilde T$ defined in (\ref{eqn:Ttilde}). Substituting simple two-forms $\Sigma(e)$ and $\bar\Sigma(e)$ into the minimal coupling ansatz (\ref{eqn:minc2}) gives
\begin{equation}
S_\zeta^\mathrm{P}[\Sigma(e),\bar\Sigma(e),\omega,\bar\omega,\zeta] = S_\zeta^\mathrm{EC}[e,\omega,\bar\omega,\zeta]\,,
\end{equation} 
from which one obtains
\begin{equation}
\frac{\delta S_\zeta^\mathrm{EC}}{\delta e^{CC'}_s} = \frac{\delta S_\zeta^\mathrm{P}}{\delta \Sigma^{PQ}_{pq}}\frac{\delta \Sigma^{PQ}_{pq}}{\delta e^{CC'}_s} +  \frac{\delta S_\zeta^\mathrm{P}}{\delta \bar\Sigma^{P'Q'}_{pq}}\frac{\delta \bar\Sigma^{P'Q'}_{pq}}{\delta e^{CC'}_s}\,.
\end{equation}
Contracting this expression with $e^{AA'}_s/\sqrt{|g|}$ and using (\ref{eqn:eesi}) gives
\begin{eqnarray}
\tilde T_{MNM'N'} &=& T^\Sigma{}^{pq}_{MP}\Sigma^P{}_{Npq}\epsilon_{M'N'}+T^{\bar\Sigma}{}^{pq}_{M'P'}\bar\Sigma^{P'}{}_{N'pq}\epsilon_{MN} \nonumber\\
&&{} -T^\Sigma{}^{pq}_{MN}\bar\Sigma_{M'N'pq}-T^{\bar\Sigma}{}^{pq}_{M'N'}\Sigma_{MNpq}\,.
\end{eqnarray}
A decomposition into components of independent symmetry, using the facts (\ref{eqn:5}), finally yields the translation prescription
\begin{eqnarray}\label{eqn:10}
\tilde T_{(MN)(M'N')} = - 2 T^\Sigma{}^{pq}_{MN}\bar\Sigma_{M'N'pq}\,, &\qquad&
\tilde T^{PP'}{}_{PP'} = 4 T^\Sigma{}^{pq}_{PQ}\Sigma^{PQ}_{pq}\,,\\
\tilde T_{MN}{}^{P'}{}_{P'} = -2 T^\Sigma{}^{pq}_{P(M}\Sigma^P{}_{N)pq}\,, &\qquad&
\tilde T^P{}_{P(M'N')} = -2 T^{\bar\Sigma}{}^{pq}_{P'(M'}\bar\Sigma^{P'}{}_{N')pq}\,.\nonumber
\end{eqnarray}

The equivalence of Einstein--Cartan and Plebanski gravity with matter minimally coupled as proposed is now simple to demonstrate. We observe that equation (\ref{eqn:6}) merely determines the Lagrange multiplier $\Psi$; it does not restrict the Weyl curvature part $\psi$ and so does not enter the discussion of equivalence. The simplicity constraint (\ref{eqn:mPs4}) forces the $\Sigma$ and $\bar\Sigma$ to be tetrad-induced. Then  (\ref{eqn:minc2}) implies $\delta S_\zeta^\mathrm{P}/\delta \omega_{ABp} = \delta S_\zeta^\mathrm{EC}/\delta \omega_{ABp}$  so that equation (\ref{eqn:mPs1}) becomes equivalent to the torsion condition (\ref{eqn:modtor}). Finally using the translation prescription (\ref{eqn:10}) in equations (\ref{eqn:7})--(\ref{eqn:9}) immediately proves their equivalence to equations (\ref{eqn:041}).

With this result we have achieved a completion of the known Plebanski description of vacuum gravity into a consistent gravity theory containing all standard tensor and spinor matter fields. 

\section{Discussion}\label{sec:disc}
Plebanski gravity is a reformulation of Einstein--Hilbert gravity in which the gravitational field is described by a basis of two-forms and by $SL(2,\mathbb{C})$-connection one-forms. It is important as a geometric starting point for loop quantum gravity and spin foam quantization. 

In this article we have developed a procedure for the minimal coupling of all common types of tensor and spinor field matter to Plebanski gravity. We have proven the consistency of this scheme by showing the equivalence of Plebanski gravity to non-vacuum Einstein--Cartan gravity. Our results are based on real (non-chiral) formulations of those theories, which is necessary for a clean comparison of the new Plebanski minimal coupling to standard real matter actions. As a substantial tool we have used the spinor technology introduced by Penrose, which was important in our proofs due to the existence of the spinor decomposition theorem. This allowed us to extract different symmetry components of the equations systematically, and to use the simplified equations to compare Plebanski and Einstein--Cartan theory.

Our results complete Plebanski gravity into a full classical gravity theory with matter. The general coupling prescription developed here is a major advance in comparison to previously known matter couplings that were suggested case by case. Our minimally coupled Plebanski matter actions are simply derived from the known Einstein--Cartan matter actions, and so do not involve any Lagrange multiplier constructions that are considered as `artificial' in~\cite{Capovilla:1991qb}. Even more importantly, we have proven the consistency of Plebanski matter couplings for the first time. 

There exist various ideas for geometric extensions of Plebanski gravity. The proposal put forward by Krasnov no longer restricts the two-forms to be induced by tetrads~\cite{Krasnov:2008fm,Krasnov:2009iy}; it is based on renormalization arguments that lead to a relaxation of the simplicity constraint. Investigations of the resulting theory promise gravitational effects such as a curvature dependent cosmological constant~\cite{Krasnov:2008fm} and a consistent behaviour in the interior regime of the Schwarzschild type solution~\cite{Krasnov:2007ky}. It is clearly desirable to find a geometric interpretation for the two-forms in case they are not simple. Besides the possible use of the Urbantke formula that merely reads out the metric information, one idea in this direction comes from the framework of area metric geometry~\cite{Punzi:2006hy,Punzi:2006nx}. Here the two-forms appear as the frames of a four-tensor field on spacetime that measures infinitesimal area elements. Instead of the local Lorentz invariance of the metric, the area metric admits an $SO(3,3)$-invariance. It would be nice to construct an extension of Plebanski gravity with this invariance becoming manifest. This might be achieved for instance by extending the algebra of the connection forms. An approach of this type has been used in a different context in~\cite{Robinson:1995qr}  in order to unify Einstein gravity with the Yang--Mills equations.

The matter coupling to extensions of Plebanski gravity remains to be investigated. The results of this article lay a systematic foundation for further research in this direction. Once interpretations of the two-form geometries are set, one should expect interesting effects beyond those that can be modelled in metric geometries.

\acknowledgments
The authors are happy to thank Claudio Dappiaggi, Manuel Hohmann, Niklas H\"ubel and Christian Pfeifer for many useful discussions. FT is grateful to the Studienstiftung des Deutschen Volkes for financial support while he was working on his Diploma thesis. MNRW gratefully acknowledges full financial support from the German Research Foundation DFG through the Emmy Noether grant WO 1447/1-1.


\begin{thebibliography}{00}
%\cite{Peldan:1993hi}
\bibitem{Peldan:1993hi}
  P.~Peldan,
  %``Actions for gravity, with generalizations: A Review,''
  Class.\ Quant.\ Grav.\  {\bf 11} (1994) 1087
  [arXiv:gr-qc/9305011].
  %%CITATION = CQGRD,11,1087;%%
  
\bibitem{Cartan}
  E.~Cartan,
  %``Sur les varietes a connexion affine et la theorie de la relativite generalisee,''
  Ann. Scient. Ec. Norm. Sup. {\bf 40} (1923) 325; {\bf 41} (1924) 1; {\bf 42} (1925) 17.  
  
\bibitem{Israel}
  W.~Israel, Differential forms in general relativity, Comm. Dublin Inst. Adv. Stud. A {\bf 19} (1970) 1.
  
%\cite{Plebanski:1977zz}
\bibitem{Plebanski:1977zz}
  J.~F.~Plebanski,
  %``On the separation of Einsteinian substructures,''
  J.\ Math.\ Phys.\  {\bf 18} (1977) 2511.
  %%CITATION = JMAPA,18,2511;%%  

%\cite{Capovilla:1991qb}
\bibitem{Capovilla:1991qb}
  R.~Capovilla, T.~Jacobson, J.~Dell and L.~Mason,
  %``Selfdual two forms and gravity,''
  Class.\ Quant.\ Grav.\  {\bf 8} (1991) 41.
  %%CITATION = CQGRD,8,41;%%  
  
%\cite{Ashtekar:1987gu}
\bibitem{Ashtekar:1987gu}
  A.~Ashtekar,
  %``New Hamiltonian Formulation of General Relativity,''
  Phys.\ Rev.\  D {\bf 36} (1987) 1587.
  %%CITATION = PHRVA,D36,1587;%%  
  
%\cite{Thiemann:2007zz}
\bibitem{Thiemann:2007zz}
  T.~Thiemann,
  Modern canonical quantum general relativity, Cambridge University Press 2007 
  [arXiv:gr-qc/0110034].
  %%CITATION = GR-QC/0110034;%%
  
%\cite{De Pietri:1998mb}
\bibitem{De Pietri:1998mb}
  R.~De Pietri and L.~Freidel,
  %``so(4) Plebanski Action and Relativistic Spin Foam Model,''
  Class.\ Quant.\ Grav.\  {\bf 16} (1999) 2187
  [arXiv:gr-qc/9804071].
  %%CITATION = CQGRD,16,2187;%%  
  
 %\cite{Freidel:2007py}
\bibitem{Freidel:2007py}
  L.~Freidel and K.~Krasnov,
  %``A New Spin Foam Model for 4d Gravity,''
  Class.\ Quant.\ Grav.\  {\bf 25} (2008) 125018
  [arXiv:0708.1595 [gr-qc]].
  %%CITATION = CQGRD,25,125018;%%
  
%\cite{Krasnov:2006du}
\bibitem{Krasnov:2006du}
  K.~Krasnov,
  %``Renormalizable Non-Metric Quantum Gravity?,''
  arXiv:hep-th/0611182.
  %%CITATION = HEP-TH/0611182;%%  
  
%\cite{Krasnov:2009pu}
\bibitem{Krasnov:2009pu}
  K.~Krasnov,
  %``Plebanski Formulation of General Relativity: A Practical Introduction,''
  arXiv:0904.0423 [gr-qc].
  %%CITATION = ARXIV:0904.0423;%%  
  
%\cite{Krasnov:2008ui}
\bibitem{Krasnov:2008ui}
  K.~Krasnov,
  %``Motion of a 'small body' in non-metric gravity,''
  Phys.\ Rev.\  D {\bf 79} (2009) 044017
  [arXiv:0812.3603 [gr-qc]].
  %%CITATION = PHRVA,D79,044017;%%
  
%\cite{Penrose:1960eq}
\bibitem{Penrose:1960eq}
  R.~Penrose,
  %``A Spinor approach to general relativity,''
  Annals Phys.\  {\bf 10} (1960) 171.
  %%CITATION = APNYA,10,171;%%
  
\bibitem{PenroseRindler}
  R.~Penrose and W.~Rindler,
  Spinors and spacetime I \& II, Cambridge University Press 1984.  
  
%\cite{Trautman:2006fp}
\bibitem{Trautman:2006fp}
  A.~Trautman,
  %``Einstein-Cartan theory,''
  arXiv:gr-qc/0606062.
  %%CITATION = GR-QC/0606062;%%
  
%\cite{Plebanski:1975wn}
\bibitem{Plebanski:1975wn}
  J.~F.~Plebanski,
  %``Some solutions of complex Einstein equations,''
  J.\ Math.\ Phys.\  {\bf 16} (1975) 2395.
  %%CITATION = JMAPA,16,2395;%%  
  
%\cite{Urbantke:1984eb}
\bibitem{Urbantke:1984eb}
  H.~Urbantke,
  %``On Integrability Properties Of SU(2) Yang-Mills Fields. I. Infinitesimal
  %Part,''
  J.\ Math.\ Phys.\  {\bf 25} (1984) 2321.
  %%CITATION = UWTHPH-1984-02;%%
  
%\cite{Krasnov:2008fm}
\bibitem{Krasnov:2008fm}
  K.~Krasnov,
  %``Plebanski gravity without the simplicity constraints,''
  Class.\ Quant.\ Grav.\  {\bf 26} (2009) 055002
  [arXiv:0811.3147 [gr-qc]].
  %%CITATION = CQGRD,26,055002;%%
  
%\cite{Krasnov:2009iy}
\bibitem{Krasnov:2009iy}
  K.~Krasnov,
  %``Gravity as BF theory plus potential,''
  Int.\ J.\ Mod.\ Phys.\  A {\bf 24} (2009) 2776
  [arXiv:0907.4064 [gr-qc]].
  %%CITATION = IMPAE,A24,2776;%%    
  
%\cite{Krasnov:2007ky}
\bibitem{Krasnov:2007ky}
  K.~Krasnov and Y.~Shtanov,
  %``Non-Metric Gravity II: Spherically Symmetric Solution, Missing Mass and
  %Redshifts of Quasars,''
  Class.\ Quant.\ Grav.\  {\bf 25} (2008) 025002
  [arXiv:0705.2047 [gr-qc]].
  %%CITATION = CQGRD,25,025002;%%  
  
%\cite{Punzi:2006hy}
\bibitem{Punzi:2006hy}
  R.~Punzi, F.~P.~Schuller and M.~N.~R.~Wohlfarth,
  %``Geometry for the accelerating universe,''
  Phys.\ Rev.\  D {\bf 76} (2007) 101501
  [arXiv:hep-th/0612133].
  %%CITATION = PHRVA,D76,101501;%
  
%\cite{Punzi:2006nx}
\bibitem{Punzi:2006nx}
  R.~Punzi, F.~P.~Schuller and M.~N.~R.~Wohlfarth,
  %``Area metric gravity and accelerating cosmology,''
  JHEP {\bf 0702} (2007) 030
  [arXiv:hep-th/0612141].
  %%CITATION = JHEPA,0702,030;%%    
  
%\cite{Robinson:1995qr}
\bibitem{Robinson:1995qr}
  D.~C.~Robinson,
  %``A Lagrangian formalism for the Einstein Yang-Mills equations,''
  J.\ Math.\ Phys.\  {\bf 36} (1995) 3733.
  %%CITATION = JMAPA,36,3733;%%  
  
\end{thebibliography}
\end{document}